\documentstyle[prd,aps]{revtex}
 
\begin{document} 
\draft
 \title{\bf   Constant mean curvature slices in
 the extended Schwarzschild solution and collapse of the lapse. Part I}
\author{ Edward Malec$^{*,**}$ and Niall \'O Murchadha$^{*,+}$}
\address{$^{*}$ ESI, A-1090 Wien, Boltzmangasse 9, Austria.}
\address{ Institute of Physics,  Jagiellonian University,
30-059  Cracow, Reymonta 4, Poland.}
\address{$^{+}$ Physics Department, University College,
Cork, Ireland.}
 \maketitle
 \begin{abstract} We give a detailed description of the
 constant mean curvature foliations in the Schwarzschild  spacetime; show
 that the lapse collapses exponentially and compute the exponent.
 \end{abstract}
 \pacs{04.20.Me, 95.30.Sf, 97.60.Lf, 98.80.Dr}
 \section{INTRODUCTION}
 \label{sec1}
 In the analysis of General Relativity
 as a Hamiltonian system \cite{MTW} one chooses a time function and
 considers the foliation of the spacetime by the slices of constant time.
 Two natural geometrical quantities arise on such spacelike three slices.
 One is the intrinsic three metric, usually $g_{ab}$, and the other is the
 extrinsic curvature $K^{ab}$, the derivative of $g_{ab}$ along the normal
 to the slice. They are related by the constraints, which in a vacuum
 spacetime read
 \begin{eqnarray} {\cal R}^{(3)}- K^{ab}K_{ab}+(\text{tr}
 K)^2&=&0\label{ham}\\ \nabla_a K^{ab}- g^{ab}\nabla_a\text{tr} K&=&0
 \label{mom} \end{eqnarray}
 where ${\cal R}^{(3)}$ is the three scalar
 curvature. Given the initial data one chooses, essentially arbitrarily,
 the lapse, $N$, and the shift, $N^i$, which determine the magnitude and
 direction of the unit time vector relative to the normal to the slice.

 One now can write the evolution equations for the intrinsic metric and
 extrinsic curvature in vacuum, e.g. \cite{York} (the reader should be 
 warned that we follow Wald \cite{wald} in our definition of the extrinsic
 curvature, not \cite{MTW}, positive $K$ means inceasing volume to the future)
 \begin{eqnarray}
 \partial_tg_{ab} &=& 2NK_{ab} + N_{a;b} + N_{b;a}\\ \partial_tK_{ab}&=&
 N_{;ab} - N\left(R_{ab} - 2K_a^dK_{bd} + K_{ab}\text{tr}K\right)  +
 K_{ab;c}N^c + K_{ac}N^c_{;b} + K_{cb}N^c_{;a}. \end{eqnarray}
  Let us
 stress that we are using the convention of signs that gives $\text{tr} K
 = + n^{\alpha}_{;\alpha}$ where $n^{\alpha}$ is the timelike unit normal
 to the slice and $\partial_t\sqrt{g} = \sqrt{g} \left(N \text{tr} K +
 N^a_{;a}\right)$. In other words, positive $\text{tr} K$ means expansion.
 It is often useful to specify the foliation, and thus the time, by
 placing a condition on the extrinsic curvature. A very popular choice is
 to demand that the  trace of the extrinsic curvature be constant on each
 slice (`CMC slicing').

 In this paper we investigate the CMC slices of the
 extended Schwarzschild solution. The manifold consists of four segments,
 each of which can be covered by the standard Schwarzschild metric
 \begin{eqnarray} ds^2&=&-\left(1 - {2m \over r}\right) dt^2+ { dr^2 \over
 1 - {2m \over r}} +r^2 \left( d\theta^2+\sin^2\theta  d\phi^2\right)
 \label{A1} \end{eqnarray}
  where $t$ is the `static' killing vector and
 $r$ is the `areal' or `Schwarzschild' radius. In the left and right zones
 $t$ is timelike and $r > 2m$ is spacelike. In the bottom zone $r$ is
 timelike and runs forward from the past singularity at $r = 0$ to $r =
 2m$. In the top zone $r$ is also timelike and runs forward from $r = 2m$
 to the future singularity, also at $r = 0$. We do not seek the most
 general CMC slices. We are looking for those CMC slices which inherit the
 underlying spherical symmetry of the given spacetime.

 There are two
 complementary ways of analysing this problem. One way is to assume that
 one is given initial data (the intrinsic metric and extrinsic curvature)
 both parts of which have the desired symmetry. It turns out that one can
 explicitly solve the constraints.  From the momentum constraint,
 Eq.(\ref{mom}), it is clear that the extrinsic curvature must be just a
 sum of the trace term and a part which is both trace and divergence free
 (`transverse-tracefree', `TT'). There exists a unique spherically symmetric
 TT tensor. Therefore the extrinsic curvature can be written down with
 just two free parameters. On substituting into the Hamiltonian
 constraint, Eq.(\ref{ham}), one discovers that this also can be solved
 explicitly.

 The alternative approach
 is to take the given spacetime and make a coordinate transformation in
 the $(t, r)$ plane only, given by $t = h(r)$, leaving the rest untouched.
 $h(r)$ is called the height function. One now imposes the
 condition that the $t' = 0$ slice be CMC. This gives a second order
 equation for the height function which can be integrated explicitly once.
 This is enough to evaluate the intrinsic metric and extrinsic curvature
 of the slice and, of course, they agree with the expressions obtained
 using the first approach.

 One then can work out how these slices fit into
 the given spacetime and construct interesting CMC foliations. One ends up
 with a first order equation for height function which cannot be
 integrated explicitly. Nevertheless one can make qualitative statements
 about the location of the slices. In this article we focus on a
 particular class of slicings where we fix the value of tr$K$ and vary the
 parameter defining the amount of the TT component in the extrinsic
 curvature. For small values of the parameter we have two foliations, one
 which runs from one infinity to the other, and one which emerges and
 returns to one of the singularities. As the parameter increases, the
 leaves of the foliations approach one another, and at a critical value of
 the parameter, they touch. For values of the parameter greater than the
 critical one the nature of the CMC slices change. They all now run
 from infinity into the singularity.

 In this article we focus on the
 behaviour of the slices as they approach the critical value. We find the
 classic `collapse of the lapse' phenomenon. Further, by looking carefully
 at the first order equation for the height function we obtain an explicit
 expression for how the lapse decays near criticality. This article draws
 very heavily on the analysis given in \cite{beig} of the collapse of the
 lapse for maximal slices of Schwarzschild.

 \section{EXTRINSIC AND INTRINSIC GEOMETRY OF CMC SLICES}
  We can generate an essentially general
 spherically symmetric slicing of the Schwarzschild solution by making a
 coordinate transformation $t' = t'(t, r), r' = r'(t, r)$. This will give
 us a spacetime metric of the form
 \begin{equation}
 ds^2 = -g_{t't'}dt'^2
 + N_{r'}dt'dr' + g_{z'z'}dr'^2 + r^2 \left( d\theta^2 + \sin^2(\theta)
 d\phi^2\right), \label{m1} \end{equation}
 We have that $g_{t't'},
 N_{r'}$, and $g_{r'r'}$ are functions of $t'$ and $r'$. The coefficient,
 $r^2$, in front of the two-metric is the original,Schwarzschild,
 coordinate $r^2$ but can be viewed as a function of $t'$ and $r'$ as
 well. We can make a further coordinate transformation of the form $r'' =
 r''(t', r')$, leaving $t'$ unchanged. This has the effect of changing the
 $r'$ coordinate within each slice but leaving the slicing unchanged. This
 kind of transformation can be used to arrange that $\nabla t' \cdot
 \nabla r'' = 0$. This is equivalent to dragging the $r'$ coordinate along
 the normal to the slicing and thus sets the shift to zero. This will give
 us a spacetime metric
 \begin{equation} ds^2 = -N^2dt'^2  +
 g_{r''r''}dr''^2 + r^2 \left( d\theta^2 + \sin^2(\theta) d\phi^2\right).
 \label{m2} \end{equation}
 On any one given slice we can arrange that $r''
 = r$, the original Schwarzschild coordinate. However, when one tries to
 propogate this condition one discovers that it is not compatible with
 vanishing shift. Therefore one can have a metric of the form (\ref{m1})
 with $r' = r$ or a metric of the form (\ref{m2}) with zero shift but not
 both. Another choice would be to set $g_{r'r'} = 1$, i.e., to choose the
 $r'$ coordinate as the proper distance along the slice. Again this is not
 compatible with vanishing shift. One advantage that the `proper distance'
 coordinate choice has over the `$r' = z$' gauge is that, so long as the
 slice remains spacelike, the proper distance gauge always remains regular
 while the `$r' = r$' choice may well have coordinate singularities.
 However, in this article we will largely stick to the metric form
 (\ref{m2}) and ignore questions such as the `best' choice of spatial
 coordinate.

 To simplify the notation we will write the line element as
 \begin{equation} ds^2=-N^2 dt^2+a dr^2+R^2 (d\theta^2+\sin^2(\theta
 )d\phi^2), \label{m3} \end{equation}
 where the written $(t, r)$ are NOT
 the original $(t, r)$ while $R$ IS the original $r$. The geometry of
 $t=$const slices is  encoded in two places. One is the dependence of $a$
 on $r$ and the other is the relationship between $R$ and $r$. This second
 piece  is contained in the mean curvature of the surfaces of constant $r$
 as embedded two-surfaces in the spatial three-geometry,
 $$ p={2 \over \sqrt{a}R} {dR\over dr}. $$

 The only nonzero (three) extrinsic curvature
 components with mixed-case indices are
 \begin{equation}
 K_r^r={\partial_ta\over 2aN}, K_{\theta }^{\theta }=K_{\phi }^{\phi }=
 {\partial_tR\over RN}={1\over 2}(\text{tr}K-K_r^r). \end{equation}
 These
 can be viewed as evolution equations for $a$ and $R$. The evolution
 equations for the extrinsic curvature can most compactly be written as
 \begin{eqnarray} \partial_t\text{tr}K &=& \nabla_i\nabla^iN - K^i_jK^j_iN
 \label{lapse}\\ \partial_t(\text{tr}K -K_r^r) &=& NR_r^{(3)r}
 -NK_i^jK^i_j+ NK_r^r\text{tr}K+{p\partial_rN \over \sqrt{a}}. \label{4}
 \end{eqnarray}

 The form of the three-dimensional Ricci curvature
 component $R_r^{(3)r}$ is
 \begin{equation}
 R^{(3)r}_r=-{\partial_r(pR)\over \sqrt{a}R}, \label{5} \end{equation}
 while the three-dimensional scalar curvature $R^{(3)}$ is
 \begin{equation} R^{(3)}=-{2\partial_r(pR) \over R\sqrt{a}}-{(pR)^2\over
 2R^2} + {2\over R^2}. \label{2a} \end{equation}
 It turns out that the
 momentum constraint can be written as
 \begin{equation} \partial_r
 (K^r_r-\text{tr}K)= -{3\over 2}p K^r_r+{1\over 2}p\text{tr}K,
 \label{mom1} \end{equation}
 and the Hamiltonian constraint is
 \begin{equation} {1\over \sqrt{a}R}\partial_r (pR)= -{3\over 4}(K^r_r)^2-
 {1\over 4} p^2+{1\over R^2} +{1\over 2}\text{tr}K K^r_r+{1\over
 4}(\text{tr}K)^2. \label{ham1} \end{equation}

 We are interested in
 finding surfaces which have $\text{tr} K$ = constant, where
 $\text{tr}K=(1/N) \partial_t\ln (\sqrt{a}R^2)$ is the fractional rate of
 change of a coordinate volume during the temporal evolution. Assume  that
 $K\equiv \text{tr}K$ is constant on a fixed Cauchy  hypersurface. Then
 the momentum constraint (\ref{mom1}) is easily solved by
 \begin{equation} K_r^r= {K\over 3} +{2C\over R^3}, \label{10a}
 \end{equation}
 where $C$ is again a constant on the chosen Cauchy
 slice. The other components of the extrinsic curvature are
 \begin{equation} K^{\theta}_{\theta} = K^{\phi}_{\phi} = {K\over 3} -
 {C\over R^3}. \label{10b} \end{equation} 
 This can be recognised as a
 combination of the trace term plus the unique spherically symmetric TT
 tensor, the terms with coefficient $C$. Therefore CMC slices of
 Schwarzschild are completely defined by the two parameters $K$ and $C$.
 The only residual freedom is the ability to drag any surface along the
 Killing vector without disturbing either the intrinsic or extrinsic
 geometry.

 The insertion of (\ref{10a}, \ref{10b}) into the Hamiltonian
 constraint leads, after some minor manipulation, to the equation
 \begin{equation} \partial_r\left[ {R\over 4}(pR)^2  - R - {C^2\over R^3}
 - {K^2\over 9}R^3\right] =0 . \label{11a} \end{equation}
 Eq.(\ref{11a}) is solved by
 \begin{equation} (pR)^2=4\left( 1 -{\beta
 \over R} + \left({KR\over 3} -{C\over R^2}\right)^2\right)  .
 \label{12a} \end{equation}
  Here $\beta $ is essentially the integration
 constant, modified by completing the  square of $K$ and $C$ related
 terms. It is easy to show that $\beta = 2m$. If `r' is replaced by the
 `areal radius, `R' then one finds $a= 4/(pR)^2$. Notice also that the
 three-dimensional line elements reads in such a case
  \begin{equation}
 ds^2_{(3)}= {4\over (pR)^2}dR^2+ R^2 (d\theta^2+\sin^2 (\theta )d\phi^2).
 \label{13a} \end{equation}

\section{THE CYLINDRICAL CMC SLICES OF THE SCHWARZSCHILD 
SPACETIME}
In the upper and lower quadrants of the Schwarzschild spacetime the Killing 
vector is spacelike and runs along the $r =$ constant surfaces.  Since 
everything is constant along the Killing vector, the trace of the trace of the extrinsic
 curvature  is preserved along these cylindrical surfaces. Therefore each $r = $ 
constant surface  is a CMC slice.

 The trace of the
 extrinsic curvature (in the upper quadrant) is given by
 \begin{equation}
 K = {2r - 3m \over \sqrt{2mr^3 - r^4}}.\label{cmcr} \end{equation}
 This
 is large and positive near $r = 2m$, zero at $r = 3m/2$ and becomes large
 and negative as $r$ becomes small.  This transforms into
 \begin{equation}
  r^4 - 2mr^3  +{(2r - 3m )^2 \over K^2} = 0.\label{cmcr1} \end{equation}

This is a quartic equation with two real roots. One lies between $r = 2m$ and 
$r = 3m/2$ and the other between $r = 3m/2$ and $r = 0$. This is clear by looking 
for the extrema of the quartic.  To find these we just differentiate to get
 \begin{equation}
 4 r^3 - 6mr^2 + 4{{2r - 3m } \over K^2} =  2(2r- 3m)(r + {2 \over K^2}) = 0.
\label{cmcr2} \end{equation}
Therefore it has only one minima (at $r = 3m/2$) and the quartic is negative there.
Hence it has two real roots, one on each side of $r = 3m/2$.  On substituting back into
Eq.(\ref{cmcr}) it is clear that the solution of Eq.(\ref{cmcr1}) with $r > 3m/2$ has
 $K > 0$ and the solution with $r <3m/2$ has $K < 0$.

In the lower quadrant things are somewhat different. 
 The trace of the
 extrinsic curvature  is now given by
 \begin{equation}
 K = {3m - 2r \over \sqrt{2mr^3 - r^4}}.\label{cmcr3} \end{equation}
 This is because, in the upper quadrant, the future is in the direction of 
decreasing $r$ while in the lower quadrant the future is in the direction of 
increasing $r$. This now 
 is large and positive near $r = 0$, zero at $r = 3m/2$ and becomes large
 and negative as $r$ approaches $2m$. We get the same quartic, Eq.(\ref{cmcr1}),
with the same roots, but now with order reversed. The root which is less than 
$3m/2$ corresponds to $K > 0$, while the other root has $K < 0$.

Given $r$, we can work out, from Eq.(\ref{cmcr}), the value of the trace of the
extrinsic curvature. We can, in fact, work out the entire extrinsic curvature
and evaluate the constant $C$ associated with these cylindrical CMC slices. In
the upper quadrant we get
\begin{equation}
C = {3mr^3 - r^4 \over 3\sqrt{2mr^3 - r^4}}.\label{cmcr4}
\end{equation}
Therefore $C > 0$ on each of these slices.

In the bottom quadrant, the extrinsic curvature picks up a minus sign. Therefore for the
cylindrical CMC slices we get
 \begin{equation}
 K = {2r - 3m \over \sqrt{2mr^3 - r^4}}; \hskip 1cm
  C = {r^4 - 3mr^3 \over 3\sqrt{2mr^3 - r^4}},\label{cmcr5} \end{equation}
  and so $C$ is negative in the lower quadrant.

 \section{THE EMBEDDING OF CMC SLICES IN THE
 SCHWARZSCHILD SPACETIME}
 In addition to the cylindrical CMC slices described in Section III, there are many other
 spherically symmetric ones. In this section, we will discuss how they run through the
 spacetime.

 The first comprehensive analysis of CMC slices
 in Schwarzschild appeared in \cite{bci}. The analysis given here closely
 follows the analysis of the related problem of maximal slices (tr$K$ = 0)
 in the Schwarzschild solution given in \cite{beig}. Let us start off with
 the Schwarzschild metric
 \begin{equation} ds^2=-\left(1 - {2m \over
 r}\right) dt^2+ { dr^2 \over 1 - {2m \over r}} +r^2 \left(
 d\theta^2+\sin^2\theta  d\phi^2\right), \end{equation}
  and look at the
 slice given by $t = h(r)$, where $h(r)$, for obvious reasons, is called
 the height function. One way of understanding the geometry of this slice
 is to make the following coordinate transformation:
 \begin{eqnarray*}
 \bar t &= t - h(r) \hskip 2cm &t = \bar t + h(\bar r)\\ \bar r &= r
 \hskip 2.9cm &r = \bar r\\ \bar \theta &= \theta \hskip 2.9cm &\theta =
 \bar \theta\\ \bar \phi &= \phi \hskip 2.9cm &\phi = \bar \phi
 \end{eqnarray*}
 where the $\bar t = 0$ surface is the slice we are
 interested in. The transformed metric becomes
 \[ \bar g_{\mu \nu}= \left(
 \matrix{ -\left(1 - {2m \over r}\right), & -h'\left(1 - {2m \over
 r}\right) & 0 & 0 \cr -h'\left(1 - {2m \over r}\right), & \left(1 - {2m
 \over r}\right)^{-1} - h'^2\left(1 - {2m \over r}\right) & 0 & 0\cr 0 & 0
 & r^2 & 0 \cr 0 & 0 & 0 & r^2\sin^2\theta \cr} \right) \] \[ \bar g^{\mu
 \nu}= \left( \matrix{ -\left(1 - {2m \over r}\right)^{-1} + h'^2\left(1 -
 {2m \over r}\right), & -h'\left(1 - {2m \over r}\right)  & 0 & 0\cr
 -h'\left(1 - {2m \over r}\right), & \left(1 - {2m \over r}\right) & 0 & 0
 \cr 0 & 0 & {1 \over r^2} & 0 \cr 0 & 0 & 0 & {1 \over r^2\sin^2\theta}
 \cr} \right) \]
 where $h' = \partial h/\partial r$. The intrinsic metric
 is given by
 \begin{equation} ds^2=\left[\left(1 - {2m \over
 r}\right)^{-1} - h'^2\left(1 - {2m \over r}\right)\right]dr^2 +r^2 \left(
 d\theta^2+\sin^2\theta  d\phi^2\right). \label{im} \end{equation}
 The lapse $N$
 of this slicing is given by
 \begin{equation} N=\left[\left(1 - {2m
 \over r}\right)^{-1} - h'^2\left(1 - {2m \over r}\right)\right]^{-{1
 \over 2}}, \label{N} \end{equation} the shift $N_a$ by
 \begin{equation} N_a=\left[-h'\left(1 - {2m \over r}\right), 0, 0
 \right], \label{N_a} \end{equation}
 and the future-pointing unit normal
 by \begin{equation} n^{\mu}={\left[\left(1 - {2m \over r}\right)^{-1} -
 h'^2\left(1 - {2m \over r}\right), h'\left(1 - {2m \over r}\right), 0, 0
 \right] \over \sqrt{\left(1 - {2m \over r}\right)^{-1} - h'^2\left(1 -
 {2m \over r}\right)}}. \end{equation}
 Given any three slice in the four
 manifold, we can drag it along by the killing vector. This will give a
 slicing where the time translation is along the killing vector. It is
 this slicing that is generated by the coordinate transformation above.
 Therefore the $N$ and $N_a$ defined by Eqns.(\ref{N}) and
 (\ref{N_a}) are nothing more than the projections of the killing vector
 perpendicular to and onto the given slice. Of course, the slicing given
 by dragging along the killing vector cannot form a foliation because the
 killing vector has a fixed point on the bifurcation sphere.

 The mean
 curvature of the $\bar t = 0$ slice is given by
 \begin{equation} K =
 n^{\mu}_{;\mu} = {1 \over \sqrt{-g}}\left(\sqrt{-g}n^{\mu}\right)_{,\mu}=
 {1 \over r^2}\partial_r\left[{r^2h'\left(1 - {2m \over r}\right) \over
 \sqrt{\left(1 - {2m \over r}\right)^{-1} - h'^2\left(1 - {2m \over
 r}\right)}}\right]. \end{equation}
 If $K$ is a constant this can be
 integrated to give
  \begin{equation} {Kr \over 3} - {C \over r^2} =
 {h'\left(1 - {2m \over r}\right) \over \sqrt{\left(1 - {2m \over
 r}\right)^{-1} - h'^2\left(1 - {2m \over r}\right)}}, \end{equation}
 where $C$ is a constant of integration. In turn, this can be manipulated
 to give
 \begin{equation} \left(1 - {2m \over r}\right)^{-1} - h'^2\left(1
 - {2m \over r}\right) = {1 \over \left(1 - {2m \over r}\right) +
 \left({Kr \over 3} - {C \over r^2}\right)^2}, \end{equation}
 and hence
 \begin{equation} h' = {{Kr \over 3} - {C \over r^2} \over \left(1 - {2m
 \over r}\right) \sqrt{\left(1 - {2m \over r}\right) + \left({Kr \over 3}
 - {C \over r^2}\right)^2}}.\label{h'} \end{equation}
 If one could integrate this one more time and find $h(r)$ in closed form,
 we would have a complete description of the slices. We cannot do so.
 Nevertheless, we can extract a significant amount of information from
 Eq.(\ref{h'}) as it stands.

 First, from the expressions Eq.(\ref{im}), Eq.(\ref{N}), and Eq.(\ref{N_a})
 it is clear that the intrinsic metric, the lapse, and the shift depend only
 on $h'$`.
Thus we get
 \begin{equation} N = {pr \over 2}= \sqrt{\left(1 - {2m \over
 r}\right) + \left({Kr \over 3} - {C \over r^2}\right)^2};\hskip 1cm  N_r =
  {{C \over r^2} - {Kr \over 3} \over 
   \sqrt{\left(1 - {2m \over r}\right) + \left({Kr \over 3}
 - {C \over r^2}\right)^2}}\label{pr1}.
 \end{equation}
 Finally, we can find the extrinsic curvature of the slice
 by using
 \begin{equation} 2N K_{ab} = {\partial g_{ab} \over
 \partial \bar t} - N_{a;b} - N_{b;a}, \end{equation}
 and we
 recover Eqns. (\ref{10a}) -- (\ref{13a}).

 When we look at Eq.(\ref{h'}), it is clear that the right hand side does
 not decay for large $r$ and thus the integral must diverge as we approach
 infinity. This is not surprising as we expect the CMC slices to go to null
 infinity.

 This agrees with
 the behaviour of the spherical CMC slices in Minkowski space. Consider the
 mass
 hyperboloid $t^2 - r^2 = 9/K^2$ in flat spacetime. If we choose the one
 which goes to future null infinity then the future-pointing timelike
 normal is $n^{\alpha} = (t, r)/\sqrt{t^2 - r^2}$, where we have to choose
 the positive root of $\sqrt{t^2 - r^2}$. We then find tr$K =
 n^{\alpha}_{;\alpha} = 3/\sqrt{t^2 - r^2} = |K| > 0$. To find out where on
 null infinity the slice ends up, we need to introduce null coordinates
 $v = (t + r)/2, u = (t - r)/2$. Using $t = r\sqrt{1 + 9/r^2K^2} \approx
 r + 9/2rK^2$ it is clear that as $v \rightarrow \infty$ $ u \approx 9/rK^2
 \rightarrow 0$. If we time translate it to $(t - t_0)^2 = r^2 + `9/K^2$
 we find $u \rightarrow t_0$. Therefore it slides up null infinity.

 If we look at Eq.(\ref{h'})
 for  large $r$ we see that $h' \approx r / \sqrt{9/K^2 + r^2}.$  The integral
  of this is $h \approx \sqrt{9/K^2 + r^2}$ which is in complete agreement with
  the flat space expression. Therefore
 the slices remain spacelike but go null infinity as $r \rightarrow
 \infty$. Further, if $K > 0$ the slices all go to future null infinity
 whereas if $K < 0$ the slices go to past null infinity.

One place we can find interesting information, without solving for $h$,
 is by looking at the expression for the mean curvature of the spherical
 two-surfaces. In particular, we
 know that
 \begin{equation} {p^2r^2\over 4} = \left(1 - {2m \over
 r}\right) + \left({Kr \over 3} - {C \over r^2}\right)^2 \ge 0. \label{pr}
 \end{equation}
 Therefore the polynomial on the right hand side of
 Eq.(\ref{pr}) must be nonnegative. Further, we know that the zeros of the
 polynomial are the points where $p = 0$ and therefore are the extrema of
 the area of the round two-spheres as embedded surfaces in the
 three-slice.

 Let us first fix some $K > 0$ and see what happens as
 we vary $C$. (The cases where $K < 0$ is remarkably
 similar.) First consider the case where $C = 0$. This is the
 so-called umbilical slice, where $K_{ab} \propto g_{ab}$. In this case
 the polynomial reduces to $1 - 2m/r + K^2r^2/9$. This is a cubic equation
 with only one real root, call it $r_u$. Outside $r = r_u$ the polynomial
 is positive, inside it, it is negative. It is clear that $r_u \le 2m$ and
 that $r_u = 2m$ iff $K = 0$. Therefore we know that the umbilical slice
 with $K > 0$ starts out at future null infinity, comes in to a minimum at
 $r = r_u$ and then passes out to the other future null infinity. The
 obvious question is whether this occurs above or below the bifurcation
 sphere.

 To settle this we need to look at the optical scalars
 \cite{he,mom,gom}
 \begin{equation} \omega_+ =2 \sqrt{\left(1 - {2m
 \over r}\right) + \left({Kr \over 3} - {C \over r^2}\right)^2} +
 2\left({Kr \over 3} - {C \over r^2}\right), \label{o+} \end{equation}
 and
 \begin{equation} \omega_- =2 \sqrt{\left(1 - {2m \over r}\right) +
 \left({Kr \over 3} - {C \over r^2}\right)^2} - 2\left({Kr \over 3} - {C
 \over r^2}\right). \label{o-} \end{equation}
 These are essentially the
 null expansions in the outgoing-future and outgoing-past directions
 respectively. They are both positive in Minkowski space and in the
 exterior regimes of the Schwarzschild solution. Since the product
 satisfies $\omega_+\omega_-/4 = 1 - 2m/r$ one or other becomes negative
 in the interior quadrants of the Schwarzschild solution. It turns out
 that the upper quadrant satisfies $\omega_+ < 0, \omega_- > 0$ while in
 the lower quadrant we have that $\omega_- < 0, \omega_+ > 0$. It is clear
 that at $r = 2m$, when $K > 0$ and $C = 0$, $\omega _-$ goes negative
 while $\omega_+$ remains positive. Therefore the umbilical slice (with $K
 > 0$) must pass through the lower quadrant. Therefore it starts at future
 null infinity, comes down so as to cross the $t = 0$ axis, passes through
 the Schwarzschild throat below the bifurcation point to some minimum
 radius $r_u$ and then rises up again to the other future null infinity.

 Let us now hold $K > 0$ fixed but change $C$ so as to be slightly larger
 than zero. Now the polynomial becomes sixth order with two roots which we
 call $r_{ms}$ ($ms$ = max.-small) and $r_{ml}$ ($ml$ = min.-large). Near
 $r = 0$ the dominant term is the positive term $C^2/r^4$ so the
 polynomial starts off large and positive while then the next term is the
 negative $-2m/r$ which pulls it negative at $r = r_{ms}$ with $r_{ms}
 \approx \sqrt[3]{C^2/2m}$. We know that the polynomial must become
 positive before $r = 2m$ and the $K^2r^2/9$ term does just that at
 $r_{ml}$ with $r_{ml} \approx r_u$. If $C > 0$ then the effect of the $C$
 term is to diminish the effectiveness of the $K^2$ term so we get that
 $r_{ml} > r_u$ while if $C < 0$ we get $r_{ml} < r_u$. Therefore for $K >
 0$ and $C > 0$ (but small) we have two different regimes in which the
 polynomial is positive. One is for small $r$ which represents a CMC slice
 which starts at $r = 0$, expands out to $r_{ms}$, which is the maximum
 area, and then contracts again back to $r = 0$.

 When we look at the null
 expansions it is clear that when $C > 0$ for small $r$ we have that
 $\omega_+ < 0$, $\omega_- > 0$ so it must be in the upper quadrant. Hence
 when $C > 0$ the small $r$ slice comes out of and goes back to the future
 singularity while the other slice runs  from future null infinity to
 future null infinity and passes through the center at a slightly larger
 radius than the umbilical one. Thus it is to the future of the umbilical
 slice and crosses closer to the  bifurcation sphere. As $C$ increases
 away from zero we continue to have two CMC slices, one which comes from
 the  future singularity out to some small radius $r_{ms} \approx
 \sqrt[3]{C^2/2m}$ and the other which goes from future null infinity to
 future null infinity but will be slightly to the future of the umbilical
 slice. We find $r_{ml}$  monotically increases as $C$ does until $C =
 8Km^3/3$. For this value of $C$, it is easy to show that $r_{ml} = 2m$ so
 that this CMC slice will pass through the bifurcation point.

 Increasing $C$ acts like a time translation near infinity. From what
 happens in Minkowski space, we expect the slice to slide up along null
 infinity.

 As $C$
 increases even further $r_{ml}$ will start to decrease again while the
 CMC slice continues to move forward in time and passes through the throat
 above the bifurcation point. As $C$ increases we find that $r_{ms}$
 increases so that the CMC slicing that begins and ends at the future
 singularity moves backwards in time.  The mimimum of the polynomial rises
 up and the two roots, $r_{ml}$  and $r_{ms}$, will approach each other as
 $C$ approaches the critical value $C = C^+_*$. For this value of
 $C$  the polynomial is everywhere positive except at one point. This will
 be at a radius we call $ R^+_*$. This will satisfy
 $R^+_* >3m/2$. $R^+_*$ is nothing more than the larger of the two
 roots of Eq.(\ref{cmcr1}) and $C^+_*$ is the value of $C$ given by
 Eq.(\ref{cmcr4}). This is because the cylindrical CMC slices act as barriers
 to the non-cylindrical CMC slices.

 As $C$ approaches $C^+_*$, each of the two CMC slices will develop
 long cylinrical regions.The one from null infinity will run along, but
 just above the surface with $r = R^+_*$ while the one from $r = 0$
 will run just below. The closer to the critical value, the longer the
 cylinders.

 When $C = C^+_*$ we get a sudden change.
 Instead of having two solutions with long cylinders we have five.
 Two come from the left and right null infinity, respectively, and
 asymptote (from above) to infinite cylinders of radius $r = R^+_*$.
 Two others come, left and right, from $r = 0$ and asymptote from below
 to the same cylinders. The fifth solution is the $r = R^+_*$
 cylinder itself.

  When $C$
 exceeds the critical value we get another change. The polynomial
 becomes everywhere positive. This means that the CMC slice cannot have
 any extremum. It must run all the way from $r = 0$ to $r = \infty$. If $C
 > C^+_*$ we will have two CMC slices, one from the left future null
 infinity which runs into the future singularity and a mirror one from the
 right future null infinity.

 Starting from the umbilical slice, holding $K$ fixed, and let $C$ become
 negative, we get the opposite behaviour. The slice from null infinity to
 null infinity moves backwards in time, while a new CMC slice emerges from
 the past singularity and goes back to it. As $C$ approaches a negative
 critical value $C^-_*$, the two roots of the polynomial approach one
 another and coincide at a radius $R^-_* < 3m/2$. This is the smaller
 root of Eq.(\ref{cmcr1}) and $C^-_*$ is the value of $C$ given by
 Eq.(\ref{cmcr5}).

  We conjecture that the slicings we have
 described for fixed
 $K$ and for $C$ in the range $C^-_* < C < C^+_*$ form three
 separate foliations, one for $r < R^-_*$ near the past singularity,
 one for $r < R^+_*$ near the future singularity, and third foliation
 formed by the slices that run from one null infinity to the other. We
 further conjecture that these three foliations cover the entire extended
 spacetime.

 \section{DIFFERENTIATING THE HEIGHT FUNCTION}
 Let us consider the foliation that runs from null infinity to null infinity.
 Each slice has the same value of $K$ but $C$ spans an interval. We could use
 the value of $C$ as a label on the slices but we want to use some
 time coordinate as label. The obvious choice is the `time at infinity'. This
 is given by
 \begin{equation} \tau(C) = \int_{r_{ml}}^{\infty}{\left({Kr \over 3} - {C \over r^2}
 \right) dr \over \left(1 - {2m
 \over r} \right)\sqrt{1 - {2m \over r} + \left({Kr \over 3}
 - {C \over r^2}\right)^2}}
 = \int_{r_{ml}}^{\infty}{\left({Kr^3 \over 3} - C\right) dr
 \over \left(1 - {2m
 \over r}\right) \sqrt{r^4 - 2mr^3 + \left({Kr^3 \over 3}
 - C\right)^2}}
 .\label{h1} \end{equation}
 This has three divergences. The first is due to the $(1 -2m/r)$, which diverges
 at the horizon. This can be integrated through in the Cauchy principal value
 sense. A similar divergence arose in \cite{beig}. The second is due to the fact
 that the polynomial inside the square root vanishes at $r = r_{ml}$. This is the
 definition of $r_{ml}$. This is not a problem because the polynomial goes to zero
 linearly at $r_{ml}$. Therefore the integral is of the form $\int dx/\sqrt{x}$,
 which is regular at $x = 0$.

 The third divergence is due to the fact that the integral itself diverges as
 $r \rightarrow \infty$. This has to be because the slice goes to null infinity.
 To leading order the integral becomes
 \begin{equation}
 \tau \approx \int {r dr \over \sqrt{{9 \over K^2} + r^2}} = \sqrt{r^2
 + {9 \over K^2}}
 \end{equation}
 which is just the flat spacetime mass hyperboloid. If we want a finite time label
 on the CMC slices the obvious thing to do would be to subtract off the leading flat
 space divergent expression. Unfortunately, the difference still logarithmically
 diverges (like $2m \ln r$). If we want a finite expression it is better to subtract
 off the height function of some favored slice of the foliation itself. One
 obvious choice is to pick the umbilical slice (the $C = 0$) slice.
  Therefore a natural time label is given by
 \begin{equation} \tau(C) = \int_{r_{ml}}^{\infty}{\left({Kr^3 \over 3} - C\right)dr
 \over \left(1 - {2m \over r}\right) \sqrt{r^4 - 2mr^3 +
 \left({Kr^3 \over 3} - C \right)^2}}
 - \int_{r_u}^{\infty}{{Kr^3 \over 3}dr
 \over \left(1 - {2m \over r}\right)
 \sqrt{r^4 - 2mr^3 + \left({Kr^3 \over 3}\right)^2}}.
\label{h2} \end{equation}

This, from the argument given above, is finite for all $C < C_*$. As $C
\rightarrow C_*$, we have that $r_{ml} \rightarrow R_*$. At this point,
the two roots of the polynomial coincide, and the slope of the tangent to the
polynomial at $r = r_{ml}$ goes to zero. The integral close to $r_{ml}$ approximates
$\int a dx /\sqrt{sx} $ where $a$ is some constant and $s$
is the slope. Integrating this over some small but finite fixed interval $(0, \Delta x)$
we get a contribution to $\tau$ of $a \sqrt{\Delta x}/2\sqrt{s}$. As $s \rightarrow 0$,
this contribution becomes unboundedly large. Therefore we get `collapse of the lapse'
in the interior. The foliation moves only a finite distance at the center to reach
$r = R_*$ while the passage of `time at infinity' becomes unboundedly large.

At the critical point both $1/h'$ and the first derivative of $1/h'$ vanish 
at the throat.
 The coefficient in the exponential decay is nothing more than the second
 derivative of $1/h'$ at the critical point. This is the dominant term in
 any expansion of the time function near the critical point. The rest of this
 article is devoted to demonstrating this.

We wish to investigate the behaviour of the central lapse.
In \cite{beig} we discussed the situation where we had a foliation defined by
some time function $\tau$ with lapse $\alpha$. Say we are given a vector field
$\xi^{\mu}$. The projection of $\xi$ normal to the time slice (call it $N$) is
given by
\begin{equation}
N = \alpha \xi^{\mu}\nabla_{\mu}\tau \Rightarrow \alpha =
N\left( \xi^{\mu}\nabla_{\mu}\tau \right)^{-1}.\label{a}
\end{equation}
If we choose $\xi$ to be the Killing vector, we know what $N$ is from Eq.(\ref{N})
and we also can write
\begin{equation}
\left( \xi^{\mu}\nabla_{\mu}\tau \right)^{-1} = \left({d\tau \over dC}\right)^{-1}
\left.{dh \over dC}\right|_r \Rightarrow \alpha = \left({d\tau \over dC}\right)^{-1}
N\left.{dh \over dC}\right|_r.    \label{l1}
\end{equation}

To evaluate expression (\ref{l1}) we need to differentiate the height function
with respect to $C$. This looks to be highly unpleasant. The square root in
the denominator is promoted to 3/2 power so the integral has a term $dx/x^{3/2}$
which diverges at the origin. Further, $r_{ml}$ depends on $C$ so there will also
be an end-point variation. This will take the integrand (which is infinite)
outside the integral sign. We know $d\tau /dC$ must be finite so these infinities
must cancel. A very similar problem arose in \cite{beig} and  a
way was found around it. This essentially involved an integration by parts before
differentiating and much more malleable expressions were found. We can repeat
this trick.

We begin by defining the following function
\begin{equation}
J = -\int {\left[r^4 - 2mr^3 +
 \left({Kr^3 \over 3} - C\right)^2\right]^{{1 \over 2}} dr \over
 \left(1 - {2m \over r}\right)}.
 \label{J}
 \end{equation}
 This is constructed so that $dJ/dC = h$. Now rewrite $J$ as
 \begin{equation}
 J = -{2 \over 3}\int {{d \over dr}\left[r^4 - 2mr^3 +
 \left({Kr^3 \over 3} - C\right)^2\right]^{{3 \over 2}} dr \over
 \left(1 - {2m \over r}\right)\left(4r^3 - 6mr^2 + 2Kr^2
 \left[{Kr^3 \over 3} - C\right]\right)}.
 \end{equation}
 This now can be integrated by parts to give
 \begin{eqnarray}
 J = &-&{2 \over 3} {\left[r^4 - 2mr^3 +
 \left({Kr^3 \over 3} - C \right)^2\right]^{{3 \over 2}} \over
 \left(1 - {2m \over r}\right)\left(4r^3 - 6mr^2 + 2Kr^2
 \left[{Kr^3 \over 3} - C\right]\right)}\cr
 &+& {2 \over 3}\int \left[r^4 - 2mr^3 +
 \left({Kr^3 \over 3} - C\right)^2\right]^{{3 \over 2}}
 {d \over dr}\left[{1 \over \left(1 - {2m \over r}\right)\left(4r^3 - 6mr^2 + 2Kr^2
 \left[{Kr^3 \over 3} - C\right]\right)}\right]dr.
 \end{eqnarray}
 We need to differentiate this twice with respect to $C$ to get $dh/dC$. We will do
 this in two parts. Let us call the not-integral part $J_1$ and the
 integral $J_2$.

We get
\begin{eqnarray}
{dJ_1 \over dC} = &2&{\left[r^4 - 2mr^3 +
 \left({Kr^3 \over 3} - C\right)^2\right]^{1 \over 2}
 \left({Kr^3 \over 3} - C\right) \over
 \left(1 - {2m \over r}\right)\left(4r^3 - 6mr^2 + 2Kr^2
 \left[{Kr^3 \over 3} - C\right]\right)}\cr
 &-&{4 \over 3}{\left[r^4 - 2mr^3 +
 \left({Kr^3 \over 3} - C \right)^2\right]^{{3 \over 2}} Kr^2\over
 \left(1 - {2m \over r}\right)\left(4r^3 - 6mr^2 + 2Kr^2
 \left[{Kr^3 \over 3} - C\right]\right)^2},
 \end{eqnarray}
 \begin{eqnarray}
 {d^2J_1 \over dC^2} = -&2&{\left[r^4 - 2mr^3 +
 \left({Kr^3 \over 3} - C\right)^2\right]^{-{1 \over 2}}
 \left({Kr^3 \over 3} - C\right)^2 \over
 \left(1 - {2m \over r}\right)\left(4r^3 - 6mr^2 + 2Kr^2
 \left[{Kr^3 \over 3} - C\right]\right)}\cr
 -&2&{\left[r^4 - 2mr^3 +
 \left({Kr^3 \over 3} - C\right)^2\right]^{1 \over 2}
  \over
 \left(1 - {2m \over r}\right)\left(4r^3 - 6mr^2 + 2Kr^2
 \left[{Kr^3 \over 3} - C\right]\right)}\cr
 +&8&{\left[r^4 - 2mr^3 +
 \left({Kr^3 \over 3} - C\right)^2\right]^{1 \over 2}Kr^2
 \left({Kr^3 \over 3} - C\right) \over
 \left(1 - {2m \over r}\right)\left(4r^3 - 6mr^2 + 2Kr^2
 \left[{Kr^3 \over 3} - C\right]\right)^2}\cr
 -&{16 \over 3}&{\left[r^4 - 2mr^3 +
 \left({Kr^3 \over 3} - C\right)^2\right]^{3 \over 2}
K^2r^4 \over
 \left(1 - {2m \over r}\right)\left(4r^3 - 6mr^2 + 2Kr^2
 \left[{Kr^3 \over 3} - C\right]\right)^3}.
 \end{eqnarray}
 One interesting property of $d^2J_1/dC^2$ is that it vanishes for large $r$.
 This means that it does not contribute to $d\tau / dC$. Note also that the
 first term in $d^2J_1/dC^2$ diverges as $r \rightarrow r_{ml}$. However,
 we must remember that to compute $\alpha$ we multiply by $N$ which goes
 to zero in the matching fashion so that everything is regular. Further, only
 the first term is finite at the throat, all the other ones vanish.

 Now we can work out
 \begin{equation}
N\left.{dh \over dC}\right|_{r_{ml}} =\left.-2{
 \left({Kr^3 \over 3} - C\right)^2 \over r^2
 \left(1 - {2m \over r}\right)\left(4r^3 - 6mr^2 + 2Kr^2
 \left[{Kr^3 \over 3} - C\right]\right)}\right|_{r_{ml}},\label{NhC}
 \end{equation}
 From the definition of $r_{ml}$ as the zero of the polynomial, Eq.(\ref{pr}),
 it is clear that
 \begin{equation}
\left({Kr^3 \over 3} - C\right)^2_{r_{ml}} = \left(2mr^3 - r^4\right)_{r_{ml}};
\left({Kr^3 \over 3} - C\right)_{r_{ml}} = -\sqrt{\left(2mr^3 - r^4\right)}_{r_{ml}}.
\label{K}\end{equation}
>From Eq.(\ref{cmcr}) we have that $K = (2R_* - 3m)/\sqrt{2mR_*^3 -R_*^4}$.
 When these are substituted into Eq.(\ref{NhC}) we get
\begin{equation}
N\left.{dh \over dC}\right|_{r_{ml}} = \left.{1 \over 2r - 3m - (2R_* - 3m)
\sqrt{{2mr^3 - r^4 \over 2mR_*^3 - R_*^4}}}\right|_{r_{ml}}.\label{NhC1}
\end{equation}
A natural variable to use (as in \cite{beig}) is $\delta = r_{ml} - R_*$. We then get
\begin{equation}
N\left.{dh \over dC}\right|_{r_{ml}} \approx {1 \over \left[2 +
{(3m - 2R_*)^2 \over 2mR_* - R_*^2}\right]\delta} = {2mR_* - R_*^2 \over
(2R_*^2 - 8mR_* + 9m^2)\delta}. \label{NhC3}
\end{equation}
The polynomial 
\begin{equation}
 D = 4r^3 - 6mr^2 + 2Kr^2
 \left[{Kr^3 \over 3} - C\right] \label{D}
 \end{equation}
  in Eq.(\ref{NhC}) is the first derivative of the
 sextic polynomial of Eq.(\ref{pr}). In general it does not vanish at $r_{ml}$.
However, we can see that
\begin{equation}
D\approx 2R_*^2\left[2 +
{(3m - 2R_*)^2 \over 2mR_* - R_*^2}\right]\delta \label{D1}
\end{equation}
and thus, as expected, goes to zero as $C\rightarrow C_*$.

Now we need to look at the integral part of $J$ as this is what gives us $d\tau/dC$.
\begin{equation}
J_2 = {2 \over 3}\int \left[r^4 - 2mr^3 +
 \left({Kr^3 \over 3} - C\right)^2\right]^{{3 \over 2}}
 {-3r^2 + 7mr - {5K^2r^4 \over 6} -3m^2 + {4mK^2r^3 \over 3} +\left(Kr - mK \right)C
 \over \left[ 2r^3 - 7 mr^2 + {K^2r^5 \over 3} +6m^2r - {2mKr^4 \over 3} +
 \left(2mKr - Kr^2\right)C\right]^2}dr
 \end{equation}
\begin{eqnarray}
&{dJ_2 \over dC} = -2\int \left[r^4 - 2mr^3 +
 \left({Kr^3 \over 3} - C\right)^2\right]^{{1 \over 2}}
\left({Kr^3 \over 3} - C\right)
{-3r^2 + 7mr - {5K^2r^4 \over 6} -3m^2 + {4mK^2r^3 \over 3} +\left(Kr - mK \right)C
 \over \left[ 2r^3 - 7 mr^2 + {Kr^5 \over 3} +6m^2r - {2mKr^4 \over 3} +
 \left(2mKr - Kr^2\right)C\right]^2}dr \cr
&\hskip 1cm + {2 \over 3}\int \left[r^4 - 2mr^3 +
 \left({Kr^3 \over 3} - C\right)^2\right]^{{3 \over 2}}
 {\left(Kr - mK \right)
 \over \left[ 2r^3 - 7 mr^2 + {K^2r^5 \over 3} +6m^2r - {2mK^2r^4 \over 3} +
 \left(2mKr - Kr^2\right)C\right]^2}dr \cr
&-{4 \over 3}\int \left[r^4 - 2mr^3 +
 \left({Kr^3 \over 3} - C\right)^2\right]^{{3 \over 2}}
{\left(-3r^2 + 7mr - {5K^2r^4 \over 6} -3m^2 + {4mK^2r^3 \over 3}
+\left(Kr - mK \right)C\right)\left(2mKr - Kr^2\right) \over
\left[ 2r^3 - 7 mr^2 + {K^2r^5 \over 3} +6m^2r - {2mK^2r^4 \over 3} +
 \left(2mKr - Kr^2\right)C\right]^3}dr.
 \end{eqnarray}
\begin{eqnarray}
{d^2J_2 \over dC^2} = {d \tau \over dC} =
&+&2\int_{r_{ml}}^{\infty} \left[r^4 - 2mr^3 +
 \left({Kr^3 \over 3} - C\right)^2\right]^{-{1 \over 2}}
\left({Kr^3 \over 3} - C\right)^2\times \cr
& &\hskip 4cm {-3r^2 + 7mr - {5K^2r^4 \over 6} -3m^2 + {4mK^2r^3 \over 3} +\left(Kr - mK \right)C
 \over \left[ 2r^3 - 7 mr^2 + {Kr^5 \over 3} +6m^2r - {2mKr^4 \over 3} +
 \left(2mKr - Kr^2\right)C\right]^2}dr. \cr
 &+&  {\rm eight\ \  other\ \  terms} \label{d2J}
 \end{eqnarray}

All the nine terms in Ew.(\ref{d2J}) fall off like $1/r^3$. Therefore each of these
terms is finite. We also know that $d\tau/dC \rightarrow \infty$ as $C \rightarrow
C_*$. The term we have isolated is the term which generates this behaviour.
All the other terms remain finite. To estimate the blowup we need to understand
the behaviour of it near $r_{ml}$. It is useful to shift the origin of
coordinates to $r = r_{crit} = R_*$. Therefore we define $y = r - R_*$.
We know $K = (2R_* - 3m)/\sqrt{2mR_*^3 -R_*^4}$ and we
also write $C = C_* - \epsilon$, where $C_* =
(3mR_*^3 - R_*^4)/3\sqrt{2mR_*^3 -R_*^4}$.

We now write out the polynomial Eq.(\ref{pr}) in terms of $(R_*, m, y, z)$ to give
\begin{eqnarray}
r^4 \dots &=& {2R_*^3 - 8mR_*^2 +9m^2R_* \over 2m - R_*}y^2 +
{12R_*^2 -56R_*m + 54m^2 \over 3(2m - R_*)}y^3 +
{17R_*^2 -54mR_* + 45m^2 \over 3a(2m -R_*)}y^4
\cr &+&
{4R_*^2 -12R_*m + 9m^2 \over 9R_*^3(2m - R_*)}[6R_*y^5 + y^6] +
{2R_*^3 - 4R_*^2m \over \sqrt{2mR_* - R_*^2}}\epsilon + \epsilon^2 +
{4R_* - 6m \over 3\sqrt{2mR_* - R_*^2}}[3R_*^2y + 3R_*y^2 +y^3]\epsilon.\label{pra}
\end{eqnarray}
The polynomial begins at $y^2$ because we know that when $\epsilon = 0$ both the
polynomial itself, and its first derivative vanishes at $y = 0$.
 More generally, we know that the polynomial vanishes when $r = r_{ml}$, i.e.,
 when $y = r_{ml} - R_* = \delta$. If $\epsilon$ is small, if we are close to the critical
 value, then
 \begin{equation}
{2R_*^3 - 8mR_*^2 +9m^2R_* \over 2m - R_*}\delta^2 + {2R_*^3 - 4R_*^2m \over 
\sqrt{2mR_* - R_*^2}}\epsilon
\approx 0 \Rightarrow \epsilon \approx {2R_*^4 - 8mR_*^3 + 9m^2R_*^2 \over 
(4R_*^2m - 2R_*^3)
\sqrt{2mR_* - R_*^2}}\delta^2. \label{dz}
\end{equation}
Further, near $r = r_{ml}$, we find that the polynomial approximates
\begin{equation}
r^4 \dots = {2R_*^3 - 8mR_*^2 +9m^2R_* \over 2m - R_*}(y^2 - \delta^2).\label{rd}
\end{equation}
The poynomial $D$ of Eq.(\ref{D}) is the first derivative of Eq.(\ref{pra}). Thus 
we have
\begin{equation}
D \approx 2{2R_*^3 - 8mR_*^2 +9m^2R_* \over 2m - R_*}y +O(y^2) \label{D2}
\end{equation}
and
\begin{equation}
{dD \over dr} = 2{2R_*^3 - 8mR_*^2 +9m^2R_* \over 2m - R_*} + O(y). \label{D'}
\end{equation}
We also need to approximate the other terms in Eq.(\ref{d2J}).
The denominator equals $ (1 - 2m/r)^2D^2/4$.  When we write this out
in terms of $(R_*, m, y)$ we get
\begin{equation}
2r^3 \dots \approx \left(1 - {2m \over r}\right)^2
{2R_*^3 - 8mR_*^2 +9m^2R_* \over 2m - R_*}^2y^2 + O(y^3)
\end{equation}
The polynomial in the numerator of Eq.(\ref{d2J}) is
\begin{equation}
-{1 \over 2}{d \over dr}\left[\left(1 - {2m \over r}\right)D\right].
\end{equation}
 Therefore
\begin{equation}
-3r^2 + \dots = -{1 \over 2}(1 - {2m \over r}){2R_*^3 - 8mR_*^2 +9m^2R_* \over 2m - R_*}
+O(y) .
\end{equation}
Therefore, the term in Eq.(\ref{d2J}) which blows up as $\delta \rightarrow 0$
is dominated by
\begin{eqnarray}
{d^2J_2 \over dC^2} = {d \tau \over dC} \approx
&+&\int_{\delta}^{\infty}{\left({Kr^3 \over 3} - C\right)^2 \over{2m \over r} - 1}
\left({2m -R_* \over 2R_*^3 - 8mR_*^2 +9m^2R_*}\right)^{3 \over 2}
{1 \over y^2\sqrt{y^2 - \delta^2}}dy.
\end{eqnarray}
We know that
\begin{equation}
{\left({Kr^3 \over 3} - C\right)^2 \over{2m \over r} - 1} \approx R_*^4,
\end{equation}
and
\begin{equation}
 \int{1 \over y^2\sqrt{y^2 - \delta^2}}dy = {\sqrt{y^2 - \delta^2} \over y\delta^2}.
 \end{equation}
 Therefore we have
 \begin{equation}
 {d^2J_2 \over dC^2} = {d \tau \over dC} \approx
 \left({2m -R_* \over 2R_*^3 - 8mR_*^2 +9m^2R_*}\right)^{3 
 \over 2}{R_*^4 \over \delta^2}.
 \label{ad2J} \end{equation}
 
 \section{COLLAPSE OF THE LAPSE}
 We now have calculated the various terms necessary to compute the central lapse.
>From Eq.(\ref{NhC3}) we have
\begin{equation}
 N\left.{dh \over dC}\right|_{r_{ml}} \approx {2mR_* - R_*^2 \over
(2R_*^2 - 8mR_* + 9m^2)\delta}. \label{NhC4}
\end{equation}
>From Eq.(\ref{dz}) we have
\begin{equation}
\epsilon = C_* - C \approx {2R_*^2 - 8mR_* + 9m^2 \over (4m - 2R_*)
\sqrt{2mR_* - R_*^2}}\delta^2. \label{dz1}
\end{equation}
We also have, from Eq.(\ref{ad2J})
\begin{equation}
  {d \tau \over dC} \approx
 \left({2m -R_* \over 2R_*^3 - 8mR_*^2 +9m^2R_*}\right)^{3 \over 2}
 {R_*^4 \over \delta^2}.
 \label{ad2J1} \end{equation}
 We can differentiate Eq.(\ref{dz1}) to get
 \begin{equation}
{dC \over d\delta} \approx -{2R_*^2 - 8mR_* + 9m^2 \over (2m - R_*)
\sqrt{2ma - R_*^2}}\delta. \label{dz2}
\end{equation}
We multiply Eq.(\ref{ad2J1}) by Eq.(\ref{dz2}) to get
\begin{equation}
  {d \tau \over dC}{dC \over d\delta} = {d\tau \over d\delta} \approx
 -\left({R_*^4 \over 2R_*^2 - 8mR_* +9m^2}\right)^{1 \over 2}{1 \over \delta}.
 \label{ad2J2} \end{equation}
 Integrating Eq.(\ref{ad2J2}) gives
 \begin{equation}
 \tau = -\left({R_*^4 \over 2R_*^2 - 8mR_* +9m^2}\right)^{1 \over 2}\ln \delta + A,
 \label{td}
 \end{equation}
 where $A$ is a constant, or
 \begin{equation}
 \delta = \exp\left[-\left({2R_*^2 - 8mR_* +9m^2 \over R_*^4}\right)^{1 \over 2}
 (\tau - A)\right]. \label{d}
 \end{equation}
>From Eq.(\ref{l1}) we have
\begin{equation}
 \alpha = \left({d\tau \over dC}\right)^{-1}
N\left.{dh \over dC}\right|_r \label{ll1}
\end{equation}
Therefore to compute the central lapse we need to divide Eq.(\ref{NhC4}) by
Eq.(\ref{ad2J1}) to get
\begin{equation}
\alpha =\left({d\tau \over dC}\right)^{-1}
N\left.{dh \over dC}\right|_r
 = {2m - R_* \over
2R_*^3 - 8mR_*^2 + 9m^2R_*}
\left({2R_*^3 - 8mR_*^2 +9m^2R_* \over 2m - R_*}\right)^{3 \over 2}{\delta \over R_*^2}
=\left({2R_*^2 - 8mR_* +9m^2 \over 2mR_*^3 - R_*^4}\right)^{1 \over 2}\delta.
\label{N1}
\end{equation}
On substituting in Eq.(\ref{d}) we get
\begin{equation}
\alpha = B\exp\left[-\left({2R_*^2 - 8mR_* +9m^2 \over R_*^4}\right)^{1 \over 2}
 \tau \right], \label{N2}
 \end{equation}
 where B is a constant which equals
 \begin{equation}
 B
=\left({2R_*^2 - 8mR_* +9m^2 \over 2mR_*^3 - R_*^4}\right)^{1 \over 2}
\exp\left[\left({ 2R_*^2 - 8mR_* +9m^2 \over R_*^4}\right)^{1 \over 2}
  A\right].
  \end{equation}
  We have not evaluated $A$ so we cannot compute $B$.
  
  It is clear from Eq.(\ref{d}) that $A$ sets the zero of $\tau$. When we compute
  the collapse of the lapse for the maximal case the moment of time symmetry
  slice sets a natural zero for the time function. In the CMC case we discuss here
  we cannot use the Killing time at infinity because it is infinite. As discussed
  in the beginning of Section V we have to normalize it by setting the zero of
  time to be that of the umbilical slice, the slice which has $C = 0$. In some sense,
  this is the analogue of the moment of time symmetry slice, but at the same 
  time it is somewhat arbitrary. This arbitrariness will be reflected in the
  constant $A$. It is also quite difficult to compute. We would need to evaluate
  integral (\ref{h2}) as we approach the critical slice. The leading term should agree
  with Eq.(\ref{td}) but we also need to compute the next term, which will give us
  $A$.

  This indicates a different way of computing the exponent. Let us look at $h'$ 
  as given by Eq.(\ref{h'}), or rather,let us look at $1/h'^2$ near the critical
  point:
  \begin{equation}
 {1 \over h'^2} \approx {2R_*^2 - 8mR_* +9m^2 \over R_*^4}(y^2 - \delta^2).
 \end{equation}
 When this is substituted into Eq.(\ref{h2}) we get
 \begin{equation}
 \tau(\delta) \approx \int_{\delta}{R_*^2 \over \sqrt{2R_*^2 - 8mR_* +9m^2}}
 {dy \over \sqrt{y^2 - \delta^2}} \label{td1}
 \end{equation}
 We have
 \begin{equation}
 \int_{\delta}{dy \over \sqrt{y^2 - \delta^2}} = 
 \ln\left|y + \sqrt{y^2 - \delta^2}\right|_{\delta} = -\ln \delta.
 \end{equation}
 Therefore we reproduce Eq.(\ref{td}).

 There are many ways of deriving the critical exponent. In a companion paper,
 \cite{mom1}, we offer a very different derivation, based on an explicit
 formula for the lapse function for spherical CMC slices. The computation
 given here, and the calculation there entirely agree.
 
 The calculation here is closely modelled on the computation of the critical 
 exponent for the maximal foliation given in \cite{beig}. This allows us to 
 perform a number of internal consistency checks at various points in the 
 calculation by reducing to the $K = 0$ situation. We get agreement at each
 stage. This agreement is not trivial because the key integration by
 parts to obtain an explicitly finite derivative of $J$  differs
 in the two cases.

 \section{Foliations with non-constant K?}
 
 As is clear from the discussion, one has three parameters to play with in 
 constructing spherical CMC foliations or slicings in the Schwarzschild solution.
 One can change $K$, one can change $C$, one can drag a slice along the `timelike'
 Killing vector. Of course, ons can change more than one of these at once.
 The foliation we have focused on is one where we kept $K$ fixed, changed $C$
 and eliminated the Killing freedom by considering the slices where the minimal
 surface coincided with the $t = 0$ axis in standard coordinates. 
 
 We could consider
 any one of these slices and drag it along the Killing vector. For the slices
 which run from null infinity to null infinity, one with $|C|$ small,
  one would get a slicing which looks
 somewhat like the standard $t =$ constant slicing of Schwarzschild. It would rise up
 along one null infinity and sink down on the other while the throat ran along one
 of the $R =$ constant lines in either the upper or lower quadrant. These would not
 form a foliation, each slice crosses each other in the interior, the lapse function will
 vanish on the throat and be positive on one side and negative on the other. If we
 drag one of the slices which run from null infinity to the singularity, one with
 $|C|$ large, along the Killing vector we get a foliation, the slices do not cross,
 but each one ends on the $R = 0$ singularity. Such a foliation would cover
 the upper half of the right hand quadrant and all the upper quadrant if one picked
 $K > 0$ and $C > 0$. Other patches could be covered by choosing other options
 for $K$ and $C$.
 
 In closed cosmologies, on the other hand, we are used to CMC foliations where
 the value of $K$ changes. If the cosmology goes from a big bang to a big crunch
 we might expect to have a foliation which goes from $K = +\infty$, 
 at the big bang, through $K = 0$,
 the moment of maximum expansion, and monotonically to $K = -\infty$, the big
 crunch. It can be shown that no such foliations with varying $K$ exists in the 
 Schwarzschild solution.

A first try would be to consider the slicing where one changes $K$ while keeping $C$ fixed. Such slices always cross each other. Consider two slices, one with $K = 0, C = 0$ (this is the standard moment-of-time-symmetry, $t = 0$, slice through Schwarzschild), and the other with $K = 1, C = 0$. As we discussed previously, this slice starts at future null infinity, crosses in the middle below the bifurcation point and rises up again to the other future null infinity. This slice crosses the first slice twice. Choosing a different, fixed value of $C$ will not change this behaviour. Therefore, if we want a foliation with varying $K$, we need a nonconstant $C$.
 
 One almost has such a foliation. Consider the slicing where one changes $K$
 while simultaneously changing $C$ such that $C = 8m^3K/3$. Each of these slices
 has its throat at $R = 2m$. Each of these slices runs through the bifurcation 
 point and so they must all touch there. They do not cross, however, and this is
 their only point of contact. This slicing covers all of the left and right
 quadrants. One might hope that by slightly changing $C$, one could spread the
 slices apart along the vertical $t = 0$ axis and thus convert this slicing into
 a foliation. This cannot be done, as we show below, if we want to allow $K$ to be unboundedly large.
 
 The standard work on the way CMC slices act as barriers was written by Brill and 
 Flaherty, \cite{bf}. Among other results, they show that two slices with the same
 value of $K$ cannot touch at a single point. Further, if two CMC slices do touch at one point, the slice to the future must have a larger value of $K$ than the other one. This restriction strongly restricts the behaviour of CMC slices in the Schwarzschild solution.

Let us assume that we have a CMC foliation
 of a Schwarzschild solution which starts off at the moment of time symmetry slice
 and moves up. Consider one slice of this foliation. This slice will have positive $K = K_S$ and $C > 8m^3K/3$. This slice will have a throat with some radius $R_S$. The cylindrical slice with this given radius is also a CMC slice (with $K = K_1$, say) and touches the other CMC slice at one point, the throat. Therefore we must have that $K_S < K_1$. Since we assume we have a foliation, we must have that $R_S$ monotonically decreases and hence also $K_1$. It passes through zero at $R_S = 3m/2$. However, we expect that $K_S$ to be increasing as the foliation moves forward in time so we will eventually run into a situation where $K_S = K_1$, which Brill and Flaherty forbid. Therefore any spherical CMC foliation of Schwarzschild cannot have unboundedly large values of $K$.

 \acknowledgments EM has been partially supported by the  KBN
 grant 2 PO3B 00623.

 \end{document}